\documentclass[aps,prl,twocolumn,superscriptaddress,floatfix,10pt]{revtex4}
\usepackage{graphicx}
\usepackage{amsmath}
\usepackage{color}
\usepackage{wrapfig}
\usepackage{latexsym}
\begin{document}

\title{Simple generic picture of toughness in solid polymer blends}

\author{Debashish Mukherji}
\email[]{debashish.mukherji@ubc.ca}
\affiliation{Quantum Matter Institute, University of British Columbia, Vancouver BC V6T 1Z4, Canada}
\author{Shubham Agarwal}\thanks{Contributed equally to this work and the names are written alphabetically}
\affiliation{Department of Mechanical Engineering, University of British Columbia, Vancouver BC V6T 1Z4, Canada}
\author{Tiago Espinosa de Oliveira}\thanks{Contributed equally to this work and the names are written alphabetically}
\affiliation{Departamento de Farmacoci\^encias, Universidade Federal de Ci\^encias da Sa\'ude de Porto Alegre, Porto Alegre 90050-170, Brazil}
\author{C\'eline Ruscher}\thanks{Contributed equally to this work and the names are written alphabetically}
\affiliation{Department of Mechanical Engineering, University of British Columbia, Vancouver BC V6T 1Z4, Canada}
\author{J\"org Rottler}
\affiliation{Quantum Matter Institute, University of British Columbia, Vancouver BC V6T 1Z4, Canada}
\affiliation{Department of Physics and Astronomy, University of British Columbia, Vancouver BC V6T 1Z1, Canada}


\begin{abstract}
Toughness $\mathcal{T}$ of a brittle polymeric solid can be enhanced by blending another compatible and ductile polymer. 
While this common wisdom is generally valid, a generic picture is lacking that connects 
the atomistic details to the macroscopic non-linear mechanics. 
Using all-atom and complementary generic simulations we show how a delicate balance 
between the side group contact density of the brittle polymers $\rho_{\rm c}$ and its dilution upon adding a 
second component controls $\mathcal{T}$. 
A broad range of systems follows a universal trend in $\mathcal{T}$ with ${\rm d}\rho_{\rm c}/{\rm d}\varepsilon$, 
where $\varepsilon$ is the tensile strain. The simulation data is consistent with a simple model based on the parallel spring analogy. 
\end{abstract}

\maketitle

Polymers are widely used in designing light weight, high performance organic materials~\cite{cohen10nm,Pipe15NMat,Mueller20PPS,Mukherji20AR,PLArev} that 
find use in common household utensils to complex nano-materials~\cite{cohen10nm,Pipe15NMat,Shuai17afm,Toohey07NM,Sharifi14JMCA}. Because of 
their wide applicability, these materials are often exposed to a range of environmental conditions, such as temperature~\cite{Robbins06Pol,Vagilis11mac,tempEffect,James22CMS}, 
pressure~\cite{Vagilis11mac,Cahill11PRB}, and/or mechanical deformation~\cite{Robbins06Pol,jr01pre,Toohey07NM,PMMAPCBritpol19,he21acsmac}. 
Here, one important materials property is the ability to sustain large deformation~\cite{PLArev,andzelmpre,he21acsmac,bhagat18,CFA19}. 

Most known commodity polymers, such as poly(methyl methacrylate) (PMMA) \cite{PMMAPCBritpol19,he21acsmac}, poly(lactic acid) (PLA)~\cite{he21acsmac,PLArev}, 
and polystyrene (PS) \cite{ANTICH2006139}, are brittle in their glassy states, 
i.e., small strain-to-fracture $\varepsilon_{\rm f}$ and low yield stress $\sigma_{\rm y}\simeq 0.1$ GPa due to the weak van der Waals (vdW) 
monomer-monomer interactions, the strength of which is about $k_{\rm B}T$ at room temperature, where $k_{\rm B}$ is the Boltzmann constant~\cite{kremer1990dynamics}. On the other hand, 
$\sigma_{\rm y}$ of a polymer can be enhanced by about an order of magnitude when the monomer-monomer hydrogen bonds (H$-$bond), with an interaction strength of
about $4k_{\rm B}T$~\cite{Mukherji20AR}, dominate the materials properties, 
such as in poly(acrylamide) (PAM), poly(acrylic acid) (PAA), and poly(vinyl alcohol) (PVA). 
These H$-$bonded systems, however, can either be brittle or ductile dependent on their respective 
(macro-)molecular architectures. 
Given the above discussion, mechanical response remains restricted in the single component
polymeric systems and thus often limits their broad applications.

A simple route to tune the tensile toughness $\mathcal T = \int_0^{\infty}\sigma(\varepsilon) {\rm d}\varepsilon$ of 
a brittle organic solid is by blending in another polymer with a relatively larger ductility and 
component-wise interactions. Here, $\sigma$ and $\varepsilon$ are the tensile stress and tensile strain, respectively. 
Experimentally relevant examples include, but are not limited to, PMMA-PLA~\cite{he21acsmac,plapmmaExp}, 
PAA-PVA blends \cite{bhagat18}, and/or PMMA-PVA~\cite{pvapmmaExp}. 
These systems are assumed to be fairly miscible, while abstaining to discuss the effects of immiscibility. 

In general, $\mathcal T$ of a blend varies monotonically between the two pure phases as a function of their relative mixing ratios~\cite{bhagat18}. 
However, a set of recent experiments on PMMA-PLA blends have reported a non-monotonic variation in
$\mathcal T$ with concentration~\cite{he21acsmac}. 
This behavior was attributed to the formation of ``so called" co-continuous phase, with an estimated length scale on the order of several $\mu$m \cite{he21acsmac,plapmmaExp}. 
However, a set of relatively much smaller scale all-atom simulations have found a similar trend in $\mathcal T$ \cite{mukherji22prm}, suggesting that the formation 
of a co-continuous phase may not be a necessary criterion for the toughness enhancement in these materials. 

Traditionally, extensive research has been conducted to understand the mechanics of polymeric materials, both from the 
experimental \cite{he21acsmac,ANTICH2006139,bhagat18,PLArev} and the simulation~\cite{jr01pre,PMMAPCBritpol19,mukherji22prm,andzelmpre} communities. 
Here, however, most simulation studies usually deal with single 
component systems and also predominantly at the (mesoscopic) generic level, where all-atom details are drastically coarse-grained~\cite{kremer1990dynamics}.
While such models are extremely useful in dealing with generic polymer properties, they often (in their pristine form)
lack the atomic-level details that play a key role in materials properties, unless of course
the generic model is specifically tuned to reproduce certain properties of interest. In particular, when dealing with polymer blends having very specific macromolecular structures and
interactions \cite{bhagat18,he21acsmac,mukherji22prm}, special attention should be paid. 
Therefore, a better understanding of the monomer-level (atomistic) interaction is needed to achieve a predictive 
mechanical response of the polymer blends, which to the best of our knowledge is lacking.

\begin{figure*}[ptb]
\includegraphics[width=1.0\textwidth,angle=0]{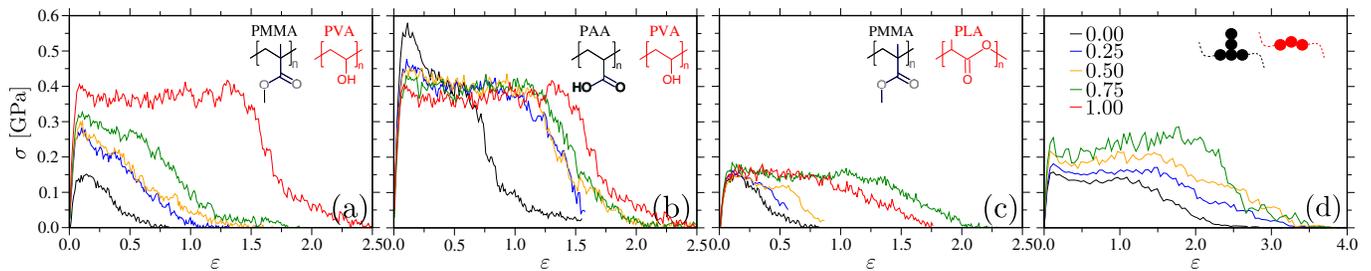}
\caption{Stress $\sigma$ as a function of strain $\varepsilon$ for three different atomistic blends, namely
poly(methyl methacrylate) (PMMA) and poly(venyl alcohol) (PVA) (panel a), poly(acitic acid) (PAA) and PVA (panel b),
PMMA and poly(lactic acid) (PLA) (panel c), and the generic model (panel d). 
The data is shown for five different PVA and PLA and four (generic) linear chain monomer mole fractions $x_{\rm i}$.
Here, $x_{\rm i} = 0.0$ corresponds to the pure brittle phase. The corresponding monomer structures are shown in the 
insets. $\sigma$ for the generic model is converted using a factor $p_{\circ}=40$ MPa~\cite{kremer1990dynamics}. 
Note that the data for PMMA-PLA blends are taken from Ref.~\cite{mukherji22prm}. 
\label{fig:ss}}
\end{figure*}

In this work, we investigate the mechanics of commodity organic solids with the goals to: 
(1) understand the effect of blending on the mechanics of brittle systems,
(2) show how the monomer (atomistic) structures play a key role in dictating the materials properties, 
(3) connect the atomistic interaction details to the macroscopic non-linear mechanics, 
and (4) propose how the behavior of many different chemically specific polymeric materials 
can be understood within one universal framework. 
To achieve the above goals, we combine all-atom and complementary generic molecular 
dynamics simulations with a simple parallel spring model.

For the all-atom simulations, we have chosen both vdW and H$-$bonded (experimentally relevant) commodity polymers 
(i.e., PMMA, PAM, PAA, PLA, and PVA) and polymer blends (i.e., PMMA-PVA, PAA-PVA, and PMMA-PLA). 
Note that we have performed the simulations for all systems in this work, aside from the PMMA-PLA trajectories which 
were obtained earlier~\cite{mukherji22prm} and reanalyzed here. 
The monomer mole fractions are varied between $0.0 \leq x_{_{\rm i}} \leq 1.0$ in steps of 0.25. 
In this study, $x_{_{\rm i}} = 0.0$ always corresponds to the pure phase of brittle polymer (BP)
and the pure second polymer (SP) at $x_{_{\rm i}} = 1.0$. 
The OPLS-AA parameters are used for PAA, PLA, and PVA~\cite{OPLS}, while the modified force-field parameters are
used for PAM~\cite{Mukherji17JCP} and PMMA~\cite{Mukherji17NC}. 
The chain length $N_{\ell} = 30$ monomers is chosen for all systems except PAM, where $N_{\ell} = 32$. 
The simulations are performed using the GROMACS molecular dynamics package~\cite{gro}. 
More system specific details are shown in the Supplementary Section S1A~\cite{epaps}. 

The well known bead-spring polymer model is used for the generic simulations~\cite{kremer1990dynamics}. 
Note that the default bead-spring model is highly ductile~\cite{jr01pre}. Therefore, inspired by the all-atom BP architectures, we have parameterized a brush-like 
polymer with short (stiff) side groups of the dimers, blended with the (ductile) linear chains. The monomer-monomer interactions are tuned to weakly mimic
the atomistic PMMA-PVA blends. The backbone chain length is taken as $N_{\ell} =30$. The generic simulations are performed using the LAMMPS package~\cite{lammps}.
A detailed discussion of the generic model is described in the Supplementary Section S1B~\cite{epaps}. The numbers that are representative of hydrocarbons are:
the units of length $d = 0.5$ nm, energy $\epsilon = 30$ meV, time $t_{\circ} = 3$ ps, and pressure $p_{\circ} = 40$ MPa~\cite{kremer1990dynamics}. Note that
for simplicity of presentation, we have converted all generic units into the real units.

In Figs.~\ref{fig:ss}(a-c) we show stress-strain behavior of
three different all-atom blends under uniaxial tensile deformation. As expected, blending in SPs drive the BP-based systems to 
a greater ductility, especially because the pure SPs are relatively more ductile than BPs.
Ideally one can argue, if SPs have their glass transition 
temperatures $T_{\rm g}$ much smaller than BP (thus remain in their rubbery phases),
the ductility or ${\mathcal T}$ may increase via rubber plasticity. 
In our case, however, all polymers have $T_{\rm g}\simeq 355-390$K (as measured in the experiments)~\cite{he21acsmac,PH}, 
while the simulations have reported $T_{\rm g}\simeq 400$K~\cite{mukherji22prm,Mukherji19PRM}.
In this context, our simulations are performed at $T =300$K and thus $T_{\rm g}-T$ are such that the individual components of 
a blend remain deep in their respective glass phases~\cite{PH,Mukherji19PRM}. This readily excludes any explanation purely 
based on $T_{\rm g}$, and it becomes apparent that a nanoscopic picture is needed to better understand the nonlinear 
mechanics of the multi-component polymeric materials. 

A comparative analysis of the monomer structures of BP 
and SP reveals that one major difference between these two sets is the presence of bulky side groups in BPs (see PAA and PMMA structures in Fig.~\ref{fig:ss}), 
while SPs are smooth (see PVA and PLA structures in Fig.~\ref{fig:ss}). 
In this context, it is known that the side groups of BPs are rather stiff
and thus form extremely rugged (instantaneous) structural corrugation along the chain contour. 
Corrugation (or more generally speaking, the breaking of translational Galilean invariance) is a necessary ``ingredient" to exert the shear 
forces upon deformation. Note also that the side groups can flip-flop normal to the chain contour 
in the polymer melt, while these structural fluctuations are frozen in the glassy state. 
In this context, it has been shown previously that the presence of stiff side chains may serve as
a necessary criterion to control the brittleness in a solid PMMA~\cite{PMMAPCBritpol19}.

\begin{figure}[ptb]
\includegraphics[width=0.37\textwidth,angle=0]{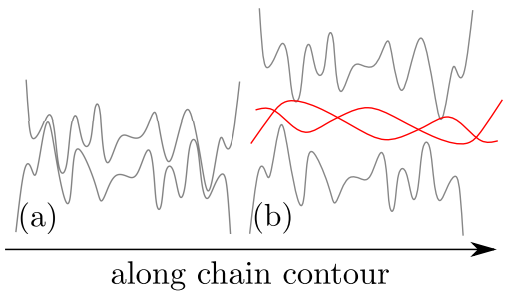}
\caption{Schematic representations of the direct contacts between the brittle polymers (BP) before (panel a) and after including
the second polymer (SP) (panel b). For the simplicity of presentation, the one-dimensional projection of the effective roughness 
(or the effective free energy landscape) of the molecules along the chain contours is shown. The grey and red lines represent (rugged) BP and (smooth) SP molecules, respectively.
\label{fig:schem}}
\end{figure}

Within the picture discussed above, the side groups of two or more neighboring chains can interlock, 
shown by the schematic in Fig.~\ref{fig:schem}(a). Here, the interlocking can be dictated by either vdW or H$-$bond
interactions. When a pure BP glass is deformed, the rigid side chain contacts break via relative motion of
the chains and thus the sliding molecules overcome the effective free energy barriers. 
Such broken contacts initiate cavitation at around the yield strain, i.e., $\varepsilon_{\rm y} \simeq 0.1-0.2$, see Fig.~\ref{fig:ss}. 
Beyond $\varepsilon_{\rm y}$, the systems go into the flow regime with growing voids, leading to their coalescence 
and ultimate fracture. 
The stronger the side chain contacts, the larger the $\sigma_{\rm y}$ values. 
In this context, we find $\sigma_{\rm y}^{\rm PMMA} < \sigma_{\rm y}^{\rm PAA} < \sigma_{\rm y}^{\rm PAM}$, see the supplementary Fig. S8~\cite{epaps}. 
This is expected because two PMMA side chains interact via weak vdW interaction, while two PAA and two PAM can form 
0.84 and 1.52 H$-$bonds per monomer, respectively.

When the SP molecules are added, they lubricate and also dilute the density of side chain contacts $\rho_{\rm c}$
between the neighboring BPs, see the schematic in Fig.~\ref{fig:schem}(b). 
This microscopic molecular arrangement increases $\varepsilon_{\rm f}$ and also hinders the
growing voids via rearrangements of SPs. 
Here, the extent of molecular ruggedness that a molecule feels, while moving in the homogeneous bulk, 
is directly related to $x_{\rm i}$ and thus has a direct implication on the polymer motion. 
The mean-square-displacement $C(t)$ data, shown in the Supplementary Figs. S2 \& S3, 
provide evidence that the molecular diffusion $D$ increases with increasing $x_{\rm i}$.
Within the linear response, the different damping contributions are linearly additive and thus the effective 
damping $\gamma \propto 1/D$. Furthermore, following $C(t)$ in the Supplementary Figs. S2 \& S3, 
it becomes apparent that the atomic level ruggedness (or corrugation) reduces with increasing $x_{\rm i}$.  
In the glassy state, we also expect the relative ruggedness across different samples 
to remain the same. 
The statement above does not aim to make a direct quantitative 
comparison between the melt and the glassy states, rather only a qualitative discussion without attempting to discuss 
the time-temperature superposition.

We note in passing that the systems investigated in this study have larger ductility than the 
experimentally reported values~\cite{he21acsmac,ANTICH2006139,bhagat18,PLArev}. This is particularly because
the quenching rates (i.e., from melt-to-glass) in the simulations are usually several orders
of magnitude faster than the corresponding experiments and thus lead to poorer annealed 
samples~\cite{statement}. Such samples usually
have residual $\sigma$ distributions that are rather broad, while well-annealed samples have narrow distribution
in $\sigma$ and thus all sites fail almost at once 
upon loading, leading to a smaller $\varepsilon_{\rm f}$ in experiments in comparison to the 
standard simulations. Here, however, it is important to note that our simulation samples are all 
prepared using an identical protocol and considering that this study aims at the relative trends, we 
believe that the data sets presented in Fig.~\ref{fig:ss} give a reasonable relative picture.

If the observation discussed above is generic, 
the trends in Figs.~\ref{fig:ss}(a-c) should also be visible in a chemically independent model by only 
incorporating the monomer structures. Therefore, we have also conducted one set of generic simulations. Here, a BP consists of a bottle brush-like 
polymer with short, stiff side groups, while a SP is represented by a standard (fully flexible) linear bead-spring chain~\cite{kremer1990dynamics}. 
In Fig.~\ref{fig:ss}(d) the generic data is presented. It can be appreciated that the trend observed in Fig.~\ref{fig:ss}(a-c) 
is also reproduced by the generic model. 

\begin{figure}[ptb]
\includegraphics[width=0.49\textwidth,angle=0]{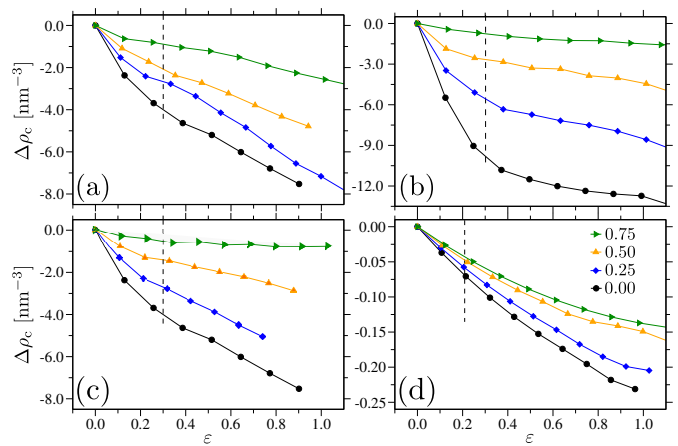}
\caption{Density of side chain contacts ${\Delta {\rho}}_{_{\rm c}} = {{\rho}}_{_{\rm c}}(\varepsilon) - {{\rho}}_{_{\rm c}}(0)$ as a function of strain $\varepsilon$ 
between the neighboring monomers of the brittle polymers in the blends of poly(methyl methacrylate) (PMMA) and poly(vinyl alcohol) (PVA) (panel a), 
poly(acrylic acid) (PAA) and PVA (panel b), PMMA and poly(lactic acid) (PLA) (panel c) and the generic model (panel d). 
For better visibility of the relative variation in the data, we have shifted all curves with 
respect to ${\rho}_{_{\rm c}}(0)$ of the unstrained system, i.e., at $\varepsilon = 0$. 
In the generic model, ${\rm d}\rho_{_{\rm c}}/{\rm d}\varepsilon$ is
scaled using the unit conversion $d = 0.5$ nm~\cite{kremer1990dynamics}.
Data for PLA-PMMA blends is derived from our earlier 
simulation trajectories in Ref.~\cite{mukherji22prm}. The vertical dashed lines show the $\varepsilon$ values at the yield points, see 
Fig.~\ref{fig:ss}.
\label{fig:contacts}}
\end{figure}

So far we have only presented a qualitative picture of the results. It will, however, be useful to draw more quantitative 
comparisons between the side chain contact density $\rho_{\rm c}(\varepsilon)$ and $\mathcal T$. Here, $\rho_{\rm c}(\varepsilon) = N_{\rm c}(\varepsilon)/v(\varepsilon)$, with  $N_{\rm c}(\varepsilon)$ and $v(\varepsilon)$ being the number of side chain contacts and the instantaneous system volume, respectively.
In the all-atom simulations, $N_{\rm c}(\varepsilon)$ is calculated when the center-of-masses of two neighboring (inter-molecular) side chains are within a minimum distance $r_{\rm min} \leq 0.65$ nm, i.e., the first peak of the radial distribution function~\cite{MM21mac,Mukherji19PRM}. In the generic model, $r_{\rm min} \leq 1.5~\sigma$~\cite{mukherji09pre}.

In Fig.~\ref{fig:contacts} we show $\rho_{\rm c}(\varepsilon)$ for the all-atom and the generic models. It can be appreciated
that: (a) $\rho_{\rm c}$ decreases with increasing $\varepsilon$, which is expected because the tensile deformation breaks 
the side chain contacts between BPs that lead to the ultimate fracture of the samples. 
(b) Increasing $x_{\rm i}$ generally dilutes $\rho_{\rm c}(\varepsilon)$ and thus leads to the weaker 
decay rates for $\varepsilon > \varepsilon_{\rm y}$ (shown by the vertical dashed lines in Fig.~\ref{fig:contacts}). 

The individual data sets in Fig.~\ref{fig:contacts} show two different ${\rm d}\rho_{\rm c}/{\rm d}\varepsilon$ regions.
The first is a relatively rapid decay for $\varepsilon_{\rm y} < 0.2$ (in all-atom) and $\varepsilon_{\rm y} < 0.1$ (in generic), 
where the ${\rm d}\rho_{\rm c}(\varepsilon)/{\rm d}\varepsilon$ behavior is dominated by the initial disruption between 
the side groups, see Fig.~\ref{fig:ss}. The second linear decay beyond $\varepsilon_{\rm y}$
occurs when the individual systems reach the flow (stationary) regime, where the voids grow and also 
SPs rearrange to control $\mathcal T$. We calculate the slopes $S = {\rm d}\rho_{\rm c}(\varepsilon)/{\rm d}\varepsilon$ in the 
stationary flow regimes, where the behavior is reasonably well described by the simple
linear form $\rho_{\rm c}(\varepsilon) = \rho_{\circ} + S\varepsilon$.

\begin{figure}[ptb]
\includegraphics[width=0.43\textwidth,angle=0]{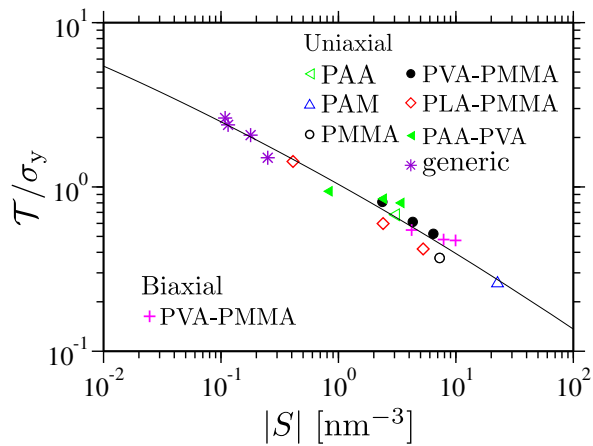}
\caption{A master curve relating the normalized toughness $\mathcal T/\sigma_{\rm y}$ 
with the change in side chain contact density $\rho_{\rm c}$ upon deformation, as
quantified by the slope $S = {\rm d}\rho_{\rm c}(\varepsilon)/{\rm d}\varepsilon$.
The data is normalized with respect to the yield stress $\sigma_{\rm y}$, calculated from 
Fig.~\ref{fig:ss}. The line is a fit to the data using Eq.~\ref{eq:emp}.
While most data sets are shown for uniaxially deformed samples, we also
present one case study in which a set of PMMA-PVA blends is biaxially deformed. 
\label{fig:uni}}
\end{figure}

Fig.~\ref{fig:uni} shows the variation of toughness-to-strength ratio $\mathcal T/\sigma_{\rm y}$ 
with $S$. It can be appreciated that a broad range of chemical specific systems and the generic model data 
fall onto one empirical master curve. 
Here, most data sets are obtained for uniaxial loading, where the mechanics is dominated by the shear force 
exerted by the sliding molecules. However, the data for the biaxial loading, where the cavitation is enhanced (see the Supplementary Figs. S6 \& S7), 
also support the same empirical behavior. 

While the data sets in Fig.~\ref{fig:uni} show a correlation across 
several samples, it is important to investigate if this behavior can be understood 
within a model by incorporating some microscopic descriptions, which is achieved using a
simple (mechanical) model. 
This is inspired by the parallel spring model analogy~\cite{network11SM}. The model specific details
are described in the Supplementary section S2 \cite{epaps}, here we only sketch the important ingredients. 
Within this model, an organic solid is modeled as a three-dimensional network 
of springs in parallel, where one side chain interlocking represents one spring 
contact with an effective spring constant $k_{i}$. Force $F$ and local displacement $\delta$ then follows
$F(\delta) = \sum_{i = 1}^{N_{\rm c}} k_i \delta$. 
If we consider that the all contacts contribute equally to the loading, i.e., $k_i \simeq k$, we get 
$F(\delta) = k N_{\rm c}(\delta) \delta$. 
Under the affine assumption, the local (microscopic) displacement $\delta$ relates to the macroscopic displacement
via $\delta = c \varepsilon L_{\circ}$. Here $L_{\circ}$ is the unstrained box dimension and $c$ is a (small) dimensionless constant. 
This assumes that each contact carry the same load and $\delta \ll \varepsilon L_{\circ}$, i.e., $c \ll 1$.
Following $\sigma$ from $F$ (see the Supplementary section S2~\cite{epaps}), we obtain
\begin{equation}
        \frac {\mathcal T}{\sigma_{\rm y}} = \frac {1}{\sigma_{\rm y}} \int_0^{\varepsilon_{\rm f}} \sigma(\varepsilon) {\rm d}\varepsilon 
    = \frac {ckL_{\circ}^2}{\sigma_{\rm y}} \int_0^{\varepsilon_{\rm f}} \rho_{\rm c}(\varepsilon) \varepsilon \left(1+\varepsilon\right){\rm d}\varepsilon.
    \label{eq:stress}
\end{equation}
Eq.~\ref{eq:stress} leads to an empirical solution,
\begin{equation}
        S \simeq \frac {\left[\frac {\mathcal T}{\sigma_{\rm y}} - \frac {ckL_{\circ}^2\rho_{\circ}}{\sigma_{\rm y}}\left\{ \frac {1}{2} \left(\frac {\mathcal T}{\sigma_{\rm y}}\right)^2 
        + \frac {1}{3} \left(\frac {\mathcal T}{\sigma_{\rm y}}\right)^3 \right\}\right]} {\frac {ckL_{\circ}^2}{\sigma_{\rm y}}\left[\frac {1}{3} \left(\frac {\mathcal T}{\sigma_{\rm y}}\right)^3 
        + \frac {1}{4} \left(\frac {\mathcal T}{\sigma_{\rm y}}\right)^4 \right]}.
    \label{eq:emp}
\end{equation}
Eq.~\ref{eq:emp} is plotted as the solid line in Fig.~\ref{fig:uni}, with $ckL_{\circ}^2\rho_{\circ}/\sigma_{\rm y} \simeq 0.01$. 
Note that the factor $k/\sigma_{\rm y}$ remains constant for a given sample.
It can be appreciated that Eq.~\ref{eq:emp} follows a reasonable agreement with the simulation 
data in Fig.~\ref{fig:uni}. This further reinforces a picture directing at a possible route towards the 
tunability in $\mathcal T/\sigma_{\rm y}$ and its links to the atomic level details. 

It will certainly require more detailed experiments to validate our scenario. 
One plausible path might be to follow the protocol presented in an earlier study of 
one of us~\cite{mukherji16SM}, where a proton nuclear magnetic resonance (NMR) setup was
used to obtain the local side chain organizations between the neighboring non-bonded monomers.
Here, we expect that the on-the-fly NMR measurements (at different $\varepsilon$) 
might give direct information about $\rho_{\rm c}$ and thus can lead to a better 
understanding of the underlying atomistic picture in these commodity 
organic solids.

In conclusion, we have performed large scale molecular dynamics simulations
to study the mechanics of solid commodity polymer blends. 
We show how a delicate balance between the (macromolecular) side chain organizations, 
resultant corrugation along the chain contour, and their interlocking controls 
the toughness $\mathcal T$ of a polymeric material. 
Our study establishes a relationship between the microscopic interactions 
and the macroscopic nonlinear mechanics, which follows 
a universal empirical relationship across a wide range of polymeric systems. 
To better understand this behavior, we have also formulated a simple
model based on the {\it springs in parallel} analogy.
The simplified picture presented here may serve as a guiding tool for the development of advanced functional 
materials with tunable and predictive mechanical properties.

D.M. thanks Carlos Marques and Kurt Kremer for stimulating discussions. 
This research was undertaken thanks, in part, to the Canada First Research Excellence Fund (CFREF), Quantum Materials and Future Technologies Program. 
D.M. further thanks ARC Sockeye facility of the University of British Columbia where 
the simulations are performed.

\bibliographystyle{ieeetr}
\bibliography{reference.bib}

\begin{thebibliography}{10}

\bibitem{cohen10nm}
M.~A. Cohen-Stuart, W.~T.~S. Huck, J.~Genzer, M.~M\"uller, C.~Ober, M.~Stamm,
  G.~B. Sukhorukov, I.~Szleifer, V.~V. Tsukruk, M.~Urban, F.~Winnik,
  S.~Zauscher, I.~Luzinov, and S.~Minko {\em Nature Materials}, vol.~9, p.~101,
  2010.

\bibitem{Pipe15NMat}
G.~Kim, D.~Lee, A.~Shanker, L.~Shao, M.~S. Kwon, Gidley, J.~Kim, and K.~P. Pipe
  {\em Nature Materials}, vol.~14, pp.~295--300, 2015.

\bibitem{Mueller20PPS}
M.~M\"uller {\em Progress in Polymer Science}, vol.~101, p.~101198, 2020.

\bibitem{Mukherji20AR}
D.~Mukherji, C.~M. Marques, and K.~Kremer {\em Annual Reviews of Condensed
  Matter Physics}, vol.~11, pp.~271--299, 2020.

\bibitem{PLArev}
X.~Zhao, H.~Hu, X.~Wang, X.~Yu, W.~Zhou, and S.~Peng {\em RSC Adv.}, vol.~10,
  pp.~13316--13368, 2020.

\bibitem{Shuai17afm}
W.~Shi, Z.~Shuai, and D.~Wang {\em Advanced Functional Materials}, vol.~27,
  p.~1702847, 2017.

\bibitem{Toohey07NM}
K.~S. Toohey, N.~R. Sottos, J.~A. Lewis, J.~S. Moore, and S.~R. White {\em
  Nature Materials}, vol.~6, pp.~581--585, 2007.

\bibitem{Sharifi14JMCA}
M.~Sharifi, C.~W. Jang, C.~F. Abrams, and G.~R. Palmese {\em Journal of
  Materials Chemistry A}, vol.~2, pp.~16071--16082, 2014.

\bibitem{Robbins06Pol}
R.~S. Hoy and M.~O. Robbins {\em Journal of Polymer Science Part B: Polymer
  Physics}, vol.~44, no.~24, pp.~3487--3500, 2006.

\bibitem{Vagilis11mac}
V.~A. Harmandaris, J.~Floudas, and K.~Kremer {\em Macromolecules}, vol.~44,
  no.~2, pp.~393--402, 2011.

\bibitem{tempEffect}
W.~Liu and W.-H. Zhai {\em Journal of Research Updates in Polymer Science},
  vol.~4, pp.~139--148, 2015.

\bibitem{James22CMS}
J.~Wu and D.~Mukherji {\em Computational Materials Science}, vol.~211,
  p.~111539, 2022.

\bibitem{Cahill11PRB}
W.-P. Hsieh, M.~D. Losego, P.~V. Braun, S.~Shenogin, P.~Keblinski, and D.~G.
  Cahill {\em Physical Review B}, vol.~83, p.~174205, 2011.

\bibitem{jr01pre}
J.~Rottler and M.~O. Robbins {\em Phys. Rev. E}, vol.~68, p.~011507, Jul 2003.

\bibitem{PMMAPCBritpol19}
K.~Fujimoto, Z.~Tang, W.~Shinoda, and S.~Okazaki {\em Polymer}, vol.~178,
  p.~121570, 2019.

\bibitem{he21acsmac}
X.~Hou, S.~Chen, J.~J. Koh, J.~Kong, Y.-W. Zhang, J.~C.~C. Yeo, H.~Chen, and
  C.~He {\em ACS Macro Letters}, vol.~10, no.~4, pp.~406--411, 2021.

\bibitem{andzelmpre}
T.~W. Rosch, J.~K. Brennan, S.~Izvekov, and J.~W. Andzelm {\em Phys. Rev. E},
  vol.~87, p.~042606, Apr 2013.

\bibitem{bhagat18}
Q.~Huang, C.~Wan, M.~Loveridge, and R.~Bhagat {\em ACS Applied Energy
  Materials}, vol.~1, no.~12, pp.~6890--6898, 2018.

\bibitem{CFA19}
M.~Huang and C.~Abrams {\em {Macromolecular Theory and Simulations}}, vol.~28,
  p.~1900030, aug 2019.

\bibitem{ANTICH2006139}
P.~Antich, A.~Vázquez, I.~Mondragon, and C.~Bernal {\em Composites Part A:
  Applied Science and Manufacturing}, vol.~37, no.~1, pp.~139--150, 2006.

\bibitem{kremer1990dynamics}
K.~Kremer and G.~S. Grest {\em The Journal of Chemical Physics}, vol.~92,
  no.~8, pp.~5057--5086, 1990.

\bibitem{plapmmaExp}
K.-P. Le, R.~Lehman, J.~Remmert, K.~Vanness, P.~M.~L. Ward, and J.~D. Idol {\em
  Journal of Biomaterials Science, Polymer Edition}, vol.~17, no.~1-2,
  pp.~121--137, 2006.

\bibitem{pvapmmaExp}
A.~{Mohammed Kadim}, A.~{Dheyaa Abdulkareem}, A.~{Jawad Kadhim alrubaie}, and
  K.~{Haneen Abass} {\em Materials Today: Proceedings}, 2021.

\bibitem{mukherji22prm}
D.~Mukherji, T.~E. de~Oliveira, C.~Ruscher, and J.~Rottler {\em Phys. Rev.
  Materials}, vol.~6, p.~025606, Feb 2022.

\bibitem{OPLS}
W.~L. Jorgensen, D.~S. Maxwell, and J.~Tirado-Rives {\em Journal of the
  American Chemical Society}, vol.~118, pp.~11225--11236, 1996.

\bibitem{Mukherji17JCP}
T.~E. de~Oliveira, D.~Mukherji, K.~Kremer, and P.~A. Netz {\em Journal of
  Chemical Physics}, vol.~146, p.~034904, 2017.

\bibitem{Mukherji17NC}
D.~Mukherji, C.~M. Marques, T.~St\"uhn, and K.~Kremer {\em Nature
  Communications}, vol.~8, p.~1374, 2017.

\bibitem{gro}
S.~Pronk, S.~Pall, R.~Schulz, P.~Larsson, P.~Bjelkmar, R.~Apostolov, M.~R.
  Shirts, J.~C. Smith, P.~M. Kasson, D.~van~der Spoel, B.~Hess, and E.~Lindahl
  {\em Bioinformatics}, vol.~29, pp.~845--854, 2013.

\bibitem{epaps}
``Electronic supplementary material. document number to be included by the
  editor.,''

\bibitem{lammps}
S.~Plimpton {\em Journal of computational physics}, vol.~117, no.~1, pp.~1--19,
  1995.

\bibitem{PH}
J.~Brandrup, E.~H. Immergut, and E.~A. Grulke, ``Polymer handbook, 2 volumes
  set, 4th edition,'' 2003.

\bibitem{Mukherji19PRM}
C.~Ruscher, J.~Rottler, C.~E. Boott, M.~J. MacLachlan, and D.~Mukherji {\em
  Physical Review Materials}, vol.~3, p.~125604, 2019.

\bibitem{statement}
 Quenching rates in simulations are usually significantly higher than the
  experiments due to the obvious computational limitations, especially when
  dealing with the all-atom trajectories of reasonably big system sizes. Of
  course, one can use particle swapping protocols to better anneal a system,
  which is, however, beyond the scope of our work.

\bibitem{MM21mac}
L.~Pigard, D.~Mukherji, J.~Rottler, and M.~Müller {\em Macromolecules},
  vol.~54, no.~23, pp.~10969--10983, 2021.

\bibitem{mukherji09pre}
D.~Mukherji and C.~F. Abrams {\em Phys. Rev. E}, vol.~79, p.~061802, Jun 2009.

\bibitem{network11SM}
R.~C. Picu {\em Soft Matter}, vol.~7, pp.~6768--6785, 2011.

\bibitem{mukherji16SM}
D.~Mukherji, M.~Wagner, M.~D. Watson, S.~Winzen, T.~E. de~Oliveira, C.~M.
  Marques, and K.~Kremer {\em Soft Matter}, vol.~12, pp.~7995--8003, 2016.

\end{thebibliography}

\end{document}